\documentclass{elsart}
\usepackage{graphics}
\usepackage{amssymb}

\begin{document}

\begin{frontmatter}

\title{Propagation of guided cold atoms}

\author{T. Lahaye, P. Cren, C. Roos and D. Gu\'ery-Odelin}
\address{Laboratoire Kastler Brossel$^*$, Ecole Normale Sup\'erieure}
\address{24, Rue Lhomond, F-75231 Paris Cedex 05, France}

\begin{abstract}
In this article we focus on the propagation of a beam of particles
guided by a transversely confining potential. We consider
different regimes. In the classical regime, we describe the beam
by means of a set of hydrodynamic-like equations. We apply this
formalism in order to investigate two practical ways for
increasing the collision rate: by using a constriction or by
tilting the guide. A high enough collision rate is indeed the most
crucial prerequisite for reaching the quantum degenerate regime by
means of evaporative cooling. In the quantum regime, we study the
propagation of bosonic atoms through a constriction in two
opposite regimes: the collision-less one and the Thomas-Fermi one.
\end{abstract}

\end{frontmatter}

\section{Introduction}
So far, Bose-Einstein condensates have only been produced in a
pulsed mode. To this respect, the achievement of a continuous beam
operating in the quantum degenerate regime remains a real
challenge. This would be the matter wave equivalent of a cw
monochromatic laser and it would allow for unprecedented
performance in terms of focalization or collimation. Recently the
MIT group has demonstrated the periodic replenishment of an
optical trap \cite{KettGuide}. This technique can therefore be
used in future experiments in combination with an appropriate
outcoupler to continuously release atoms from the trap, and hence
to obtain a truly cw atom laser.

Another possible way to achieve this goal has been studied
theoretically in \cite{Mandonnet00}. In this proposal, a
non-degenerate, but already slow and cold beam of particles, is
injected into a magnetic guide
\cite{Schmiedmayer95,Denschlag99,Goepfert99,Key00,Dekker00,Teo01,Sauer01,Hinds99}
where transverse evaporation takes place. If the elastic collision
rate is large enough, efficient evaporative cooling will lead to
quantum degeneracy after a propagation length in the guide
compatible with experimental constraints (\textit{i.e.} a few
meters). The continuous and pulsed loading of a slow and cold
atomic beam into a magnetic guide has been recently demonstrated
experimentally \cite{cren}. However, the collision rate is not
sufficient so far to initiate the evaporative cooling stage.

 In this article, we start  with summarizing the experimental techniques already developed
 for atom guiding. To achieve conditions favorable to evaporative
 cooling one has to maximize the collision rate. For that purpose, we
 describe quantitatively two possible manipulations of a
 guided thermal beam: constriction and tilting of the guide.
 We then study the role of interactions on a beam of Bose-condensed atoms
  propagating through  a constriction.

\section{Atom waveguide}
In recent years, many experimental groups have designed both
macroscopic and microscopic guides for transporting laser-cooled
atoms. In those guides, the confinement is provided by means of
conservative forces. So far the dipole and magnetic forces have
been used.

In past years, atomic transport by means of hollow optical fibers
has been achieved \cite{fiber}. In this scheme atoms are guided by
the optical dipole force. However, the heating of atoms through
intensity fluctuations and spontaneous emission may be a problem.

An alternative way consists in using magnetic guides. For alkali
atoms, magnetic forces are strong since the magnetic moment $\mu$
is of the order of the Bohr magneton $\mu_B$. As the thermal
energy of laser-cooled alkali atoms (microkelvin range) is
generally much less than the ground-state hyperfine splitting ($1$
GHz corresponds to 48 mK), we can assume that the field explored
in the guide produces a linear Zeeman shift proportional to the
hyperfine $g$-factor $g_F$. We also assume that the the spin
adiabatically follows the magnetic field, yielding an effective
potential $U=g_Fm_F\mu_B|B|$. Depending on the atom's magnetic
sublevel $m_F$ (positive or negative), atoms are attracted towards
or repelled from a magnetic extremum. Actually, Maxwell's
equations for magnetostatics only allow for the realization of a
minimum of static magnetic fields in vacuum \cite{wing}.
Therefore, only low-field seekers can be trapped with
inhomogeneous static magnetic fields. As a consequence, the atoms
are not trapped in the energetically lowest state. This effect and
the three-body recombination are responsible for the metastability
of the confined gas.

Quadrupole guides can be produced by means of four wires or tubes
equally spaced on a cylinder of radius $R$ and carrying currents
$+I$ and $-I$ alternatively. The magnetic field is quadrupolar and
is well approximated by the linear form: $B(r_\bot)=b'r_\bot$
where $r_\bot=(x^2+y^2)^{1/2}$ is the distance from the $z$ axis
and $b'\propto I/R^2$ the strength of the magnetic gradient. In
this configuration, atoms with low angular momentum are not stable
against spin flips. To avoid this loss mechanism, one superimposes
a bias field $B_0$ along the axis of the guide. Note that in this
case the field strength no longer varies linearly with radius but
is given by
\begin{equation}
B(r_\bot)=\sqrt{B_0^2+b'^2r_\bot^2}\simeq B_0 +
\frac{b'^2r_\bot^2}{2B_0}.
\end{equation}
The Taylor expansion around the center yields a quadratic
variation of the field strength. The potential experienced by the
atoms is therefore harmonic provided that their temperature $T$
fulfills the condition $k_BT\ll \mu B_0$. A constriction in the
 potential can be created by reducing the spacing between
the wires of the guide and/or decreasing the longitudinal bias
field.

\section{Increasing the collision rate of a classical gas}

  As mentioned in the introduction, the elastic collision rate $\Gamma_{\rm coll}$
 in the sample must be high enough at the entrance of magnetic guide to
start an efficient evaporative cooling \cite{Mandonnet00}. We
study in this section two ways of increasing $\Gamma_{\rm coll}$
in the classical regime for a beam of particles propagating
through the guide. Both methods ensure the conservation of phase
space density. The first one consists in a constriction of the
transverse confining potential, while the second relies on gravity
to slow down the beam by tilting the guide with respect to the
horizontal plane.

\subsection{Constriction}

\subsubsection{Single particle}
\label{onep}

Before studying the physics of a beam in the hydrodynamic regime,
it is worthwhile  to examine the one-particle motion in a
constricted guide, in order to gain physical insight. We therefore
write the equation of motion along $z$ for a particle with mass
$m$ in the power law transverse potential
$U(x,y,z)=m\lambda(z)r_\bot^\gamma$ where $\lambda(z)$ denotes the
$z$-dependent strength of the transverse confinement,
\begin{equation}
\frac{\d^2z}{\d t^2}=-r_\bot^{\gamma}\frac{\d\lambda}{\d z}\,.
\label{Eqzpoint}
\end{equation}
Equation (\ref{Eqzpoint}) shows that the constriction induces a
transfer of longitudinal energy into the transverse motion. For a
smooth constriction, the motion is adiabatic \cite{goldstein} and
consequently the quantity $E_\bot\lambda^{-2/(2+\gamma)}$ is
constant, where $E_\bot$ denotes the transverse energy. As the
total energy $E=E_\bot+m\dot{z}^2/2$ is also conserved, the
decrease in longitudinal velocity can easily be related to the
increase in $\lambda$. We notice that the effect of the
constriction is more dramatic for particles having a high
transverse energy, and is absent for a particle propagating along
the axis.

\subsubsection{Hydrodynamic regime}
We now assume that there are enough collisions to ensure local
thermodynamical equilibrium. From Appendix \ref{B} we deduce the
equations governing the beam
 temperature $T(z,t)$, linear density $n(z,t)$ and mean longitudinal velocity $u(z,t)$:
\begin{eqnarray}
&& \frac{\partial n}{\partial t} +   \frac{\partial}{\partial z}
[n u]=0,\nonumber \\
&& \frac{\partial}{\partial t} [n u]  +  \frac{\partial}{\partial
z }\bigg[\frac{n k_B T}{m}+n u^2\bigg]
+\frac{2n k_B T}{m\gamma\lambda}\frac{\d \lambda}{\d z} =0, \nonumber \\
&& \frac{\partial}{\partial t}\,
\bigg[\bigg(\frac{3}{2}+\frac{2}{\gamma}\bigg)n k_B T+\frac{mn
u^2}{2}\bigg] + \frac{\partial}{\partial
z}\bigg[\bigg(\frac{5}{2}+\frac{2}{\gamma}\bigg)n u k_B T+\frac{mn
u^3}{2}\bigg] =0\,\,. \nonumber
\end{eqnarray}
We now study the stationary regime. The previous equations become
\begin{eqnarray}
&& nu =\phi\,\,, \label{EqLambda} \\
&& \frac{\d}{\d z}\,\bigg[\frac{k_B T}{u}+mu\bigg] +\frac{2k_B
T}{\gamma u\lambda}\,\frac{\d \lambda}{\d z} =0\,\,,\label{EqdOm} \\
&& \bigg(\frac{5}{2}+\frac{2}{\gamma}\bigg)k_B T+\frac{mu^2}{2}
=\mu\,\,,\label{EqMu}
\end{eqnarray}
where $\phi$ and $\mu$ are constants. Eq. (\ref{EqLambda})
corresponds to the conservation of the number of particles, Eq.
(\ref{EqdOm}) of momentum along $z$ and Eq. (\ref{EqMu}) of
enthalpy. After some algebra one obtains the relation between the
strength of the potential
$\hat{\lambda}\equiv\lambda(z)/\lambda(0)$ normalized to its
initial value
 and the reduced
temperature $\hat{T}\equiv T(z)/T(0)$:
\begin{equation}
\hat{\lambda}=\hat{T}^{\,\gamma(\delta-2)/4}
\bigg(1-\frac{\delta}{\eta_0}(\hat{T}-1)\bigg)^{\gamma/4}\,\,,\label{EqT}
\end{equation}
where we have introduced the parameters $\delta\equiv 5+4/\gamma$
and $\eta_0=mu(0)^2/(k_BT(0))$ which compares the initial kinetic
and thermal energies. For typical experimental values \cite{cren},
one has $\eta_0\gg 1$ and (\ref{EqT}) yields
\begin{equation}
\hat{T}\simeq\hat{\lambda}^{4/(\gamma(\delta-2))}.
\end{equation}
As the normalized collision rate $\hat{\Gamma}\equiv \Gamma_{\rm
coll}(z)/\Gamma_{\rm coll}(0)$ increases as $\hat{\Gamma}\sim
\hat{T}^2$, one deduces the scaling law
$\hat{\Gamma}\sim\hat{\lambda}^{8/(\gamma(\delta-2))}\,,$ which
shows that moderate constrictions   (for example by a factor of 4)
can increase the collision rate by a large amount (nearly one
order of magnitude for an harmonic transverse confinement).

\subsection{Tilted guide}

\begin{figure}
\begin{center}
\includegraphics{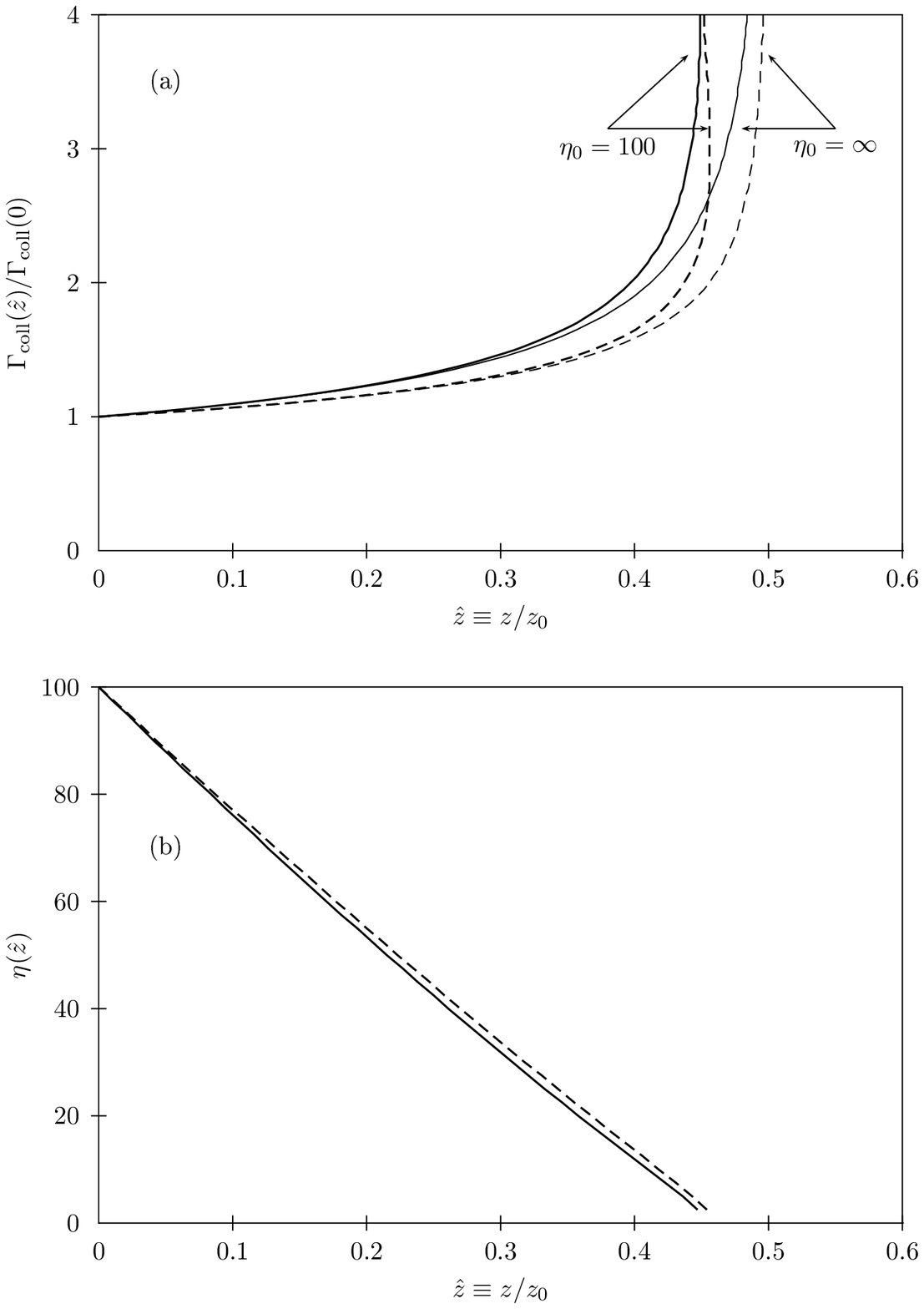}
\end{center}
\caption{ (a) Collision rate normalized at its initial value
$\hat{\Gamma}(\hat z)=\Gamma_{\rm coll}(\hat z)/\Gamma_{\rm
coll}(0)$ as a function of the reduced longitudinal coordinate
$\hat z$ for a tilted guide, for
 $\eta_0=100$ and $\eta_0=\infty$.
The dashed curves correspond to a linear transverse confinement
($\gamma=1$) and the solid lines to a harmonic transverse
confinement ($\gamma=2$). (b) Parameter $\eta\equiv mu^2(\hat
z)/(k_BT(\hat z))$ as a function of $\hat z$, for an initial value
of 100, for linear (dashed line) and harmonic (solid line)
transverse confinement.} \label{fig1}
\end{figure}

We now consider that the guide axis $z$ is tilted by an angle
$\alpha$ with respect to the  horizontal plane. The strength
 $\lambda$ of the transverse confining potential
 is assumed constant. By tilting the guide, the mean velocity of
 the beam decreases while the linear
 density and the temperature increase, resulting in an increase of
 the collision rate.
 Following the same procedure as the one developed
 in Appendix \ref{B}, we get, in the stationary regime, the following equations~:
\begin{eqnarray}
&& nu =\phi\,\,,  \\
&& \frac{\d}{\d z}\,\bigg[mu+\frac{k_B T}{u}\bigg]
+\frac{mg\sin{\alpha}}{u}
=0\,\,, \\
&& \left(\frac{2}{\gamma}+\frac{5}{2}\right)k_BT+\frac{mu^2}{2}
+mgz\sin{\alpha}=\mu\,\,,
\end{eqnarray}
where $g$ is the acceleration of gravity. This set of equations
can be solved analytically, one deduces the relation between the
reduced temperature $\hat{T}$ and the reduced velocity of the beam
$\hat{u}=u(z)/u(0)$: $\hat{T}=u^{2/(2-\delta)}$. We also establish
the relation between the collision rate and the coordinate $\hat z
\equiv z/z_0$ with $z_0=u_0^2/(g\sin\alpha)$, and the link between
the parameter $\hat{\eta}=mu^2/(\eta_0k_BT)=\hat{u}^2/\hat{T}$ and
$\hat z$:
\begin{eqnarray}
\hat{z}(\hat{\Gamma}) & = &
\frac{\delta+\eta_0}{2\eta_0}-\frac{\delta}{2\eta_0}
\hat{\Gamma}^{1/2}-\frac{1}{2}\hat{\Gamma}^{(2-\delta)/2}
\nonumber \\
\hat{z}(\hat{\eta}) & = &
\frac{\delta+\eta_0}{2\eta_0}-\frac{\delta}{2\eta_0}
\hat{\eta}^{1/(1-\delta)}-\frac{1}{2}\hat{\eta}^{(2-\delta)/(1-\delta)}.
\nonumber
\end{eqnarray}
We recall that for a particle with initial velocity $u_0$ moving
in this potential, the turning point corresponds to $\hat z=1/2$.
Fig. \ref{fig1} (a) shows the gain in the collision rate as a
function of $\hat z$ for 2 values of the parameter $\eta_0=100$
and $\eta_0=\infty$ and for an harmonic ($\gamma=2$) and a linear
($\gamma=1$) confinement. Along the propagation through the tilted
guide, the temperature $T$ increases and the mean velocity $u$
decreases resulting in a strong reduction of the parameter
$\eta(z)$ as a function of $z$ (see Fig. \ref{fig1} (b)).

 In this section, we have
described the propagation of a beam of classical particles through
a constriction and a tilted guide by means of a set of
hydrodynamic-like equations. We have used this formalism to
investigate two practical ways for increasing the collision rate.
Indeed those results are valid for a beam of bosonic particles. We
recall that a high collision rate is the most crucial requirement
for reaching the quantum degenerate regime by means of evaporative
cooling \cite{Mandonnet00}. The remainder of the paper is devoted
to the quantum degenerate regime. We focus on the propagation
through a constriction, and analyze the role of interactions.

\section{Constriction of a quantum beam}
For atomic waveguide geometries with very small transverse
dimensions as now experimentally demonstrated
\cite{Schmiedmayer95,Denschlag99,Goepfert99,Key00,Dekker00,Teo01,Sauer01,Hinds99},
the quantum nature of atoms starts to influence the dynamics. This
problem has first been investigated in the context of
quantized-conductance in a two-dimensional electron gas
\cite{constriction}. In those models, the conducting properties of
a nanoscopic constriction are investigated in the free-electron
approximation with hard-walls boundary conditions. In this
section, we address the corresponding problem for a beam of
Bose-condensed atoms confined in the transverse direction. In the
following, we restrict our study only to harmonic confinement for
the sake of simplicity. As for the classical case of the previous
section, we model the constriction by a dependence of the
transverse frequency with the longitudinal coordinate. The
potential experienced by the atoms is therefore of the form
$U(x,y,z)=m\omega(z)^2[x^2+y^2]/2$. We begin with the case of an
ideal beam, and then consider the role played by interactions for
the propagation through a constriction.

\subsection{The ideal Bose-condensed beam}

In this paragraph we study the propagation of an ideal
Bose-condensed beam. For this purpose, we solve the stationary
Schr\"odinger equation by means of the adiabatic basis (the
harmonic oscillator's one with an angular frequency that depends
on the longitudinal coordinate $z$). Details of this lengthy but
straightforward calculation can be found in the Appendix \ref{A}.
If $k$ denotes the incoming longitudinal wave vector, the local
wave vector $k_{0,0}$
  for atoms in the transverse ground state
   within the adiabatic approximation reads (see Eq. (\ref{B2})):
\begin{equation}
k^2_{0,0}=k^2-\frac{1}{\sigma^2},\nonumber
\end{equation}
$\sigma$ being the local harmonic oscillator length
$\sigma(z)=(\hbar/m\omega(z))^{1/2}$.   The physical picture is
quite clear: the constriction induces an augmentation in the
zero-point energy for the transverse degrees of freedom. As the
total energy of the atom is conserved, the longitudinal kinetic
energy decreases. Actually, one recovers, in the quantum domain,
the transfer of kinetic energy from longitudinal to transverse
degrees of freedom as already pointed out for the classical
dynamics. However, the constriction modifies the propagation even
for particles in the transverse ground state because of the finite
size of the transverse wave function.

\subsection{The role of interactions}

We investigate in this paragraph the role of interactions on the
propagation through a constriction of a beam of interacting
Bose-condensed atoms. The main difficulty arises from the
non-linearity of the Schr\"odinger equation due to the mean field
interaction term. The strength of the interaction is proportional
to the scattering length $a$ \cite{rmpsandro}. By assuming that
the transverse dimension of the cloud has the local equilibrium
shape adapted to the local number of particle per unit length, the
problem becomes one-dimensional and one can introduce a solution
for the wave function of the form $\psi ({\bf
r},t)=f(z,t)g(x,y;n)$ in the variational procedure
\cite{pethick98}, where $n$ is the local density of particles per
unit length. By writing the longitudinal part of the wave function
$f$ in the phase and modulus form $f\equiv \sqrt{n}e^{i\varphi}$,
we introduce the longitudinal velocity field defined by $u\equiv
(\hbar/m)\partial \varphi/\partial z$.

After some algebra (see Appendix \ref{C}), one gets the stationary
hydrodynamic equations for the propagation of a beam of
interacting Bose-condensed particles:
\begin{eqnarray}
&& nu=\phi \label{cons} \\ &&
\tilde{\mu}+\frac{1}{2}mu^2-\frac{\hbar^2}{2m\sqrt{n}}\frac{\partial^2\sqrt{n}}{\partial
z^2 }=c, \label{euler}
\end{eqnarray}
where $\tilde{\mu}$ is an effective chemical potential, and $\phi$
and $c$ two constants depending on the initial conditions. Eq.
(\ref{cons}) reflects the conservation of the flux while Eq.
(\ref{euler}) accounts for the conservation of energy.

As a first example, from Eq. (\ref{cons}) and (\ref{euler}) we
recover the results of the previous section for a perfect gas for
which $a=0$ and $\tilde{\mu}=\hbar\omega$ in a smooth constriction
\emph{i.e.} such that the last term of the l.h.s of (\ref{euler})
is negligible. In the following, we apply this set of equations in
the weak-coupling limit ($na\ll 1$) and in the Thomas-Fermi limit
($na\gg 1$). For those two limits, the value of $\tilde{\mu}$
differs (see Appendix \ref{C}). It is convenient to introduce the
dimensionless notations: $\hat{u}\equiv u(z)/u(0)$, $\hat{z}\equiv
z/\sigma(0)$, $\hat{\omega}\equiv \omega(z)/\omega(0)$,
$\alpha\equiv mu^2(0)/(\hbar\omega(0))$ and $\beta\equiv
\sqrt{na}=\sqrt{a\phi/u(0)}$. The combination of Eq. (\ref{cons})
and Eq. (\ref{euler}) leads to the relation between the reduced
velocity $\hat{u}$ and the rate of compression $\hat{\omega}$ as a
function of the dimensionless parameters.

In the weak-coupling limit that corresponds to the low density
regime ($\beta \ll 1$) we find:
\begin{equation}
\frac{\d^2\hat{u}}{\d\hat{z}^2}=\frac{3}{2\hat{u}}\left (
\frac{\d\hat{u}}{\d\hat{z}} \right
)^2-2\alpha\hat{u}\,(\hat{u}^2-1)+4\hat{u}\,(1-\hat{\omega}) +
8\hat{u}\beta^2(1-\hat{\omega}/\hat{u}) \label{EqAlpha}
\end{equation}
while in the opposite regime, commonly called the Thomas-Fermi
regime, we have
\begin{equation}
\frac{\d^2\hat{u}}{\d\hat{z}^2}=\frac{3}{2\hat{u}}\left (
\frac{\d\hat{u}}{\d\hat{z}} \right
)^2-2\alpha\hat{u}\,(\hat{u}^2-1)+8\beta\,(\hat{u}-\hat{\omega}\sqrt{\hat{u}})\,.
\label{EqBeta}
\end{equation}

\begin{figure}
\begin{center}
\includegraphics{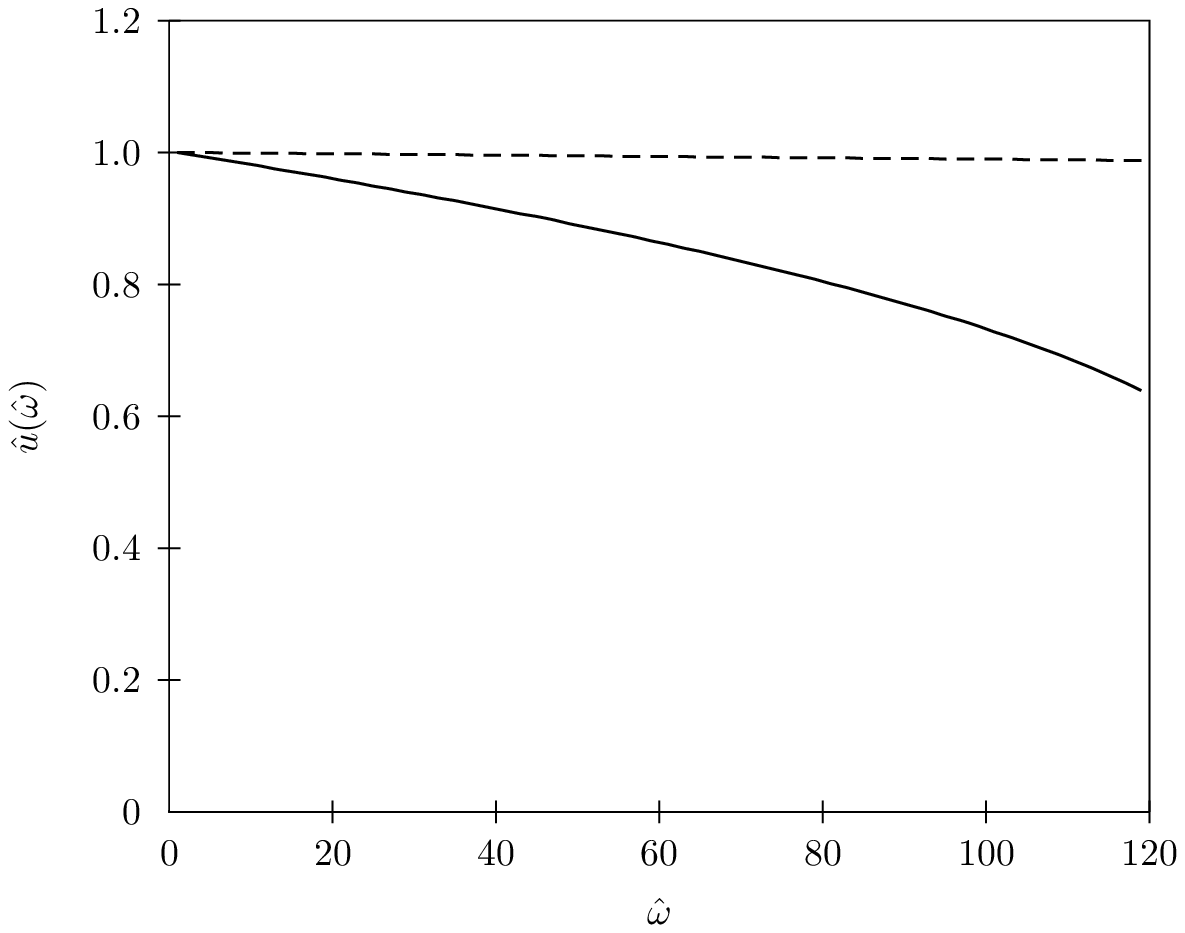}
\end{center}
\caption{Longitudinal velocity $\hat u\equiv u/u_0$ as a function
of the constriction ratio $\hat\omega$ for a quantum beam in the
weak-coupling regime (dashed line, $\beta=0.1$) and in the
Thomas-Fermi regime (solid line, $\beta=10$) for $\alpha=10^4$ and
$\gamma=10^{-4}$. In the Thomas-Fermi regime, the repulsive
interactions lead to a strong coupling between longitudinal and
transverse degrees of freedom resulting in a larger reduction of
the beam velocity through the constriction.} \label{fig2}
\end{figure}

In the following, we consider a linear increase of the transverse
confining frequency: $\hat\omega=1+\gamma\hat z$. The results of
the numerical integration of Eq. (\ref{EqAlpha}) and
(\ref{EqBeta}) are plotted on Fig. \ref{fig2}, for the following
values of the dimensionless parameters:
$\alpha=10^4,\,\gamma=10^{-4}$, $\beta$ having the value $10$ in
the Thomas-Fermi regime, and $0.1$ in the weak-coupling regime.
The large constriction ratios implied by $\gamma=10^{-4}$ can be
achieved on microchips \cite{microchips}. For a given compression
rate, the reduction of the velocity for the propagation through a
constriction is dramatically magnified in the interacting case. It
reflects the fact that the ground state wave function has a much
broader transverse profile because of repulsive interactions
($\sqrt{na}\gg 1$). As already emphasized in \S \ref{onep}, the
larger the transverse size of the beam, the higher the coupling
between longitudinal and transverse degrees of freedom. We point
out that this magnification of the coupling between degrees of
freedom induced by interactions may have to be taken into account
for interferometers based on guided matter waves \cite{erika}.

\section{Conclusion}
In this article, we have investigated quantitatively two possible
ways (by compression and tilting of the guide) of increasing the
collision rate of a thermal guided beam while keeping the phase
space density constant . Those results are important from an
experimental point of view since as soon as the collision rate is
high enough, one can implement evaporative cooling as suggested in
\cite{Mandonnet00} to reach the quantum degeneracy regime. Those
results have been obtained under the hydrodynamic formalism. In
future work, we intend to the study the case where the collision
rate is not sufficient to ensure the local equilibrium requirement
for hydrodynamics. We have also carried out the calculation for
the propagation of a beam of bosons through a constriction in the
degenerate regime in presence or not of interactions between
particles. Strictly speaking, those calculations are valid at
$T=0$. Actually, the propagation of a beam of bosons through a
constriction below the critical temperature can be regarded as a
two-fluid problem. To address this problem completely, one would
have to take into account the interaction between the thermal and
the condensed part.

\section*{Acknowledgements}
We thank Jean Dalibard and Johnny Vogels for a careful reading of
the manuscript. C. F. Roos acknowledges support from the European
Union (contract HPMFCT-2000-00478). This work was supported by the
Bureau National de la M\'etrologie, the D\'el\'egation
G\'en\'erale de l'Armement, the Centre National de la Recherche
Scientifique and the R\'egion Ile de France.

$^*$ Unit\'e de Recherche de l'Ecole normale sup\'erieure et de
l'Universit\'e Pierre et Marie Curie, associ\'ee au CNRS.

\appendix
\section{Theoretical framework for guided beams in classical regime}
\label{B}

We consider a gas of particles with mass $m$ confined by a static,
velocity-independent potential $U({\bf r})$. In this appendix we
derive the classical equations for a beam of particles confined in
a cylindrical geometry. For that purpose, we start with the
classical Boltzmann equation:
\begin{equation}
\frac{\partial f}{\partial t}+{\bf v}\cdot\frac{\partial
f}{\partial {\bf r}}-\frac{1}{m}\frac{\partial U}{\partial {\bf
r}}\cdot\frac{\partial f}{\partial {\bf v}} = I_{\rm coll}[f],
\label{ebq}
\end{equation}
where $f({\bf r}, {\bf v},t)$ is the distribution function and
$I_{\rm coll}[f]$ is collision term. The density $\rho$ is
obtained from the distribution by integration over velocity:
\begin{equation}
\rho({\bf r},t)  =   \int \d^3v f({\bf r}, {\bf v},t).
\end{equation}

\subsection{Reduction of dimensionality}
We present the formulation of conservation equations for
cylindrical geometries. We choose the longitudinal axis along $z$.
In order to get hydrodynamic-like equations, we assume that the
distribution function $f$ is given by a local Maxwell-Boltzmann
distribution of the form:
\begin{equation}
f({\bf r}, {\bf v},t)\propto\exp\left\{ -\frac{m({\bf
v}-u(z,t){\bf e}_z)^2}{2k_BT(z,t)} -\frac{U(x,y,z)}{k_BT(z,t)}\,.
\right\}\label{Ansatz}
\end{equation}
In the ansatz (\ref{Ansatz}) we assume that the velocity field is
purely longitudinal, and that all macroscopic quantities depend
only on $z$ \cite{hdrq}. We introduce the linear density $n$
defined by:
\begin{equation}
n(z,t) =  \int \d x\d y\d^3v f({\bf r}, {\bf v},t).
\end{equation}
The conservation of the number of particles is obtained by taking
the average of (\ref{ebq}) over the transverse coordinates and the
velocity components: $\int [$\ref{ebq}$]\d x \d y \d^3v$:
\begin{equation}
\frac{\partial n}{\partial t} +  \frac{\partial}{\partial
z}\bigg(n \langle v_z\rangle\bigg) = 0,  \label{eqcont}
\end{equation}
where $\langle \cdots \rangle$ denotes the following normalized
average
$$
\langle A \rangle \equiv \frac{1}{n(z)}\int \d x\d y \d^3v A
f({\bf r}, {\bf v},t)\,.
$$
 The axial momentum current equation is
deduced from the average of the on-axis component of the velocity
$\int [$\ref{ebq}$]v_z\d x \d y \d^3v$:
\begin{equation}
\frac{\partial }{\partial t}\bigg(n\langle v_z\rangle\bigg) +
\frac{\partial}{\partial z}\bigg(n \langle v^2_z\rangle\bigg)+
\frac{n}{m}\left\langle\frac{\partial U }{\partial z}\right\rangle
= 0. \label{eqimpul}
\end{equation}

The conservation of energy is obtained from the average of the
one-particle hamiltonian $H({\bf r},{\bf v})=mv^2/2+U({\bf r})$,
$\int [$\ref{ebq}$]H({\bf r},{\bf v})\d x \d y \d^3v$
\begin{equation}
\frac{\partial }{\partial t}\bigg(n\langle H\rangle\bigg) +
\frac{\partial}{\partial z}\bigg(n\langle v_z\rangle\langle
H\rangle\bigg)+\frac{m}{2}\frac{\partial}{\partial z}\bigg(n
(\langle v^3_z\rangle-\langle v_z\rangle\langle
v^2_z\rangle)\bigg)=0 \label{eqenerg}
\end{equation}

\subsection{Application to a power law potential for a stationary flow}

We consider in this section that the transverse confining
potential is given by a power law of exponent $\gamma$ with a
$z$-dependent strength $\lambda(z)$:
\begin{equation}
 U(x,y,z)=m\lambda(z) r_\bot^\gamma
 \label{powerlawpot}
\end{equation}
where   $ r_\bot=(x^2+y^2)^{1/2}$ denotes the radial distance from
the guide axis.

Furthermore, we define the mean velocity of the beam by $u
\equiv\langle v_z \rangle$, and calculate with the ansatz
(\ref{Ansatz}) the following useful quantities: $\langle v^2_z
\rangle = u^2+k_BT/m$ and $\langle v^3_z \rangle = u^3+3uk_BT/m$.
One also readily calculates the average values $\langle U\rangle =
2k_B T/\gamma$, $\langle H\rangle = (2/\gamma+3/2)k_BT+mu^2/2$.
The one-dimensional hydrodynamic equations (\ref{eqcont}) to
(\ref{eqenerg}) then simplify considerably. In the stationary
regime that we shall consider hereafter, they can be recast in the
form:
\begin{eqnarray}
&& nu =\phi ,\nonumber \\
&& \frac{\d}{\d z}\,\bigg[\frac{k_B}{u} T+mu\bigg] +\frac{2k_B
T}{\gamma\lambda u}\,\frac{\d \lambda}{\d z}
=0\,\,,\label{eqimpbis}\nonumber \\
&& \left ( \frac{2}{\gamma}+\frac{5}{2} \right )k_B
T+\frac{mu^2}{2} =\mu\,\,. \label{eqmubis}
\end{eqnarray}

An interesting feature of this set of equations is the
conservation of phase-space density $D\equiv\rho\lambda_{dB}^3$,
where $\lambda_{dB}=h/\sqrt{2\pi m k_BT}$ is the thermal de
Broglie wavelength and $\rho$ the 3D particle density. Indeed, for
the power-law potential (\ref{powerlawpot}), $D$ scales with the
four parameters $u$, $T$, $\lambda$ and $\gamma$ as: $D\sim
\lambda^{2/\gamma}u^{-1}T^{-2/\gamma-3/2}$. One can easily check
that the set of equations (\ref{eqmubis}) implies that
\[
\frac{\d D}{\d z}=0.
\]
The collision rate $\Gamma_{\rm coll}$ scales as $\rho T^{1/2}$,
thus the ratio  $\hat{\Gamma}= \Gamma_{\rm coll}(z)/\Gamma_{\rm
coll}(0)$ scales as $(T(z)/T(0))^{2}$.

\section{Compression of an ideal quantum gas}
\label{A}  In this appendix, we address the problem of the
propagation of a quantum gas in a harmonic waveguide with a
constriction. The potential experienced by the atoms reads:
$U(x,y,z)=m\omega^2(z)(x^2+y^2)/2$, $\omega(z)$ being the
$z$-dependent angular frequency. We introduce the harmonic
oscillator length $\sigma(z)=(\hbar/m\omega(z))^{1/2}$,   and
denote its first and second derivatives with respect to $z$ by
$\sigma'$ and $\sigma''$. For a theoretical description, we resort
to the study of the stationary solutions of the Schr\"odinger
equation \cite{constriction}. A general wave function can be
expanded on the adiabatic basis as $|\Psi \rangle=\sum_{n_x,n_y}
\alpha_{n_x,n_y}(z)|n_x(z)\rangle\otimes|n_y(z)\rangle$, where $\{
|n_i(z)\rangle \}$ denotes the local 1D harmonic basis for the
$i$-direction with a $z$-dependence   in angular frequency.
Inserting $|\Psi \rangle$ into the Schr\"odinger equation with
stationary energy $E\equiv \hbar^2k^2/2m$ and projecting on the
mode $|m_x(z)\rangle\otimes|m_y(z)\rangle$ yields the following
set of coupled linear equations:
\begin{eqnarray}
&&\frac{\d^2\alpha_{m_x,m_y}}{\d
z^2}+k^2_{m_x,m_y}\alpha_{m_x,m_y}
 =
- \sum_{n_x\neq m_x} \bigg(2\frac{\d\alpha_{n_x,m_y}}{\d z}A_{m_x,n_x}+\alpha_{n_x,m_y}B_{m_x,n_x}\bigg) \nonumber\\
 && -\sum_{n_y\neq m_y}\bigg(2\frac{\d\alpha_{m_x,n_y}}{\d
 z}A_{m_y,n_y}+\alpha_{m_x,n_y}B_{m_y,n_y}\bigg) \nonumber\\ &&
 + 2\sum_{n_x\neq m_x \atop n_y\neq
m_y}\alpha_{n_x,n_y}A_{m_x,n_x}A_{m_y,n_y},\label{alpham}
\end{eqnarray}

where we have defined   the matrix elements $A_{m_i,n_i}=\langle
m_i|\partial/\partial z|n_i\rangle$, $B_{m_i,n_i}=\langle
m_i|\partial^2/\partial z^2|n_i\rangle$, and the local wave number
$k_{m_x,m_y}$
\begin{eqnarray}
k^2_{m_x,m_y}=k^2-\frac{\sigma'^2}{2\sigma^2}(2+m_x+m_y+m_x^2+m_y^2)-
\frac{1+2m_x+2m_y}{\sigma^2}.\label{B2}
\end{eqnarray}
The matrix elements can be readily calculated:
\begin{eqnarray}
A_{m_i,n_i} & = & \sigma' \langle m_i| a_i^{\dagger
2}-a_i^2|n_i\rangle
/2\sigma \nonumber \\
 B_{m_i,n_i} & = & \frac{1}{4\sigma^2}\bigg(\sigma'^2 \langle m_i|
[a_i^{\dagger 2}-a_i^2]^2|n_i\rangle
-2(\sigma\sigma''-\sigma'^2)\langle m_i| a_i^{\dagger
2}-a_i^2|n_i\rangle\bigg),\nonumber
\end{eqnarray}
where $a_i^\dagger$ and $a_i$ are the creation and annihilation
operators for the considered harmonic basis.

Eq. (\ref{alpham}) shows that each mode experiences a different
potential barrier. In particular, $k_{m_x,m_y}(z)$ decreases in
for a propagation through a constriction since the ground state
energy increases in this case. From the above analysis we can
conclude that an incident wave initially in the transverse ground
state remains in this state if the adiabatic approximation
criterium is fulfilled, namely $|\sigma'|\ll 1$. The adiabatic
approximation fails around the turning point if the constriction
is such that the incident wave can be reflected. We note that if
the condition of adiabatic approximation is not fulfilled, Eq.
(\ref{alpham}) allows one to evaluate the importance of
inter-level scattering.

\section{Effective one-dimensional hydrodynamic equation in the quantum regime}
\label{C} In this appendix, we derive the effective 1D
hydrodynamic equations for a beam of Bose-condensed interacting
atoms propagating through a guide. For this purpose, we recall the
main steps of their derivations as presented in \cite{pethick98}
by means of the standard variational procedure. This approach is
valid under the dilute-gas assumption \cite{rmpsandro}, namely
$\bar{\rho} a^3 \ll 1$ where $a$ is the scattering length and
$\bar{\rho}$ the mean density.
\subsection{Hydrodynamic equations}
The action ${\mathcal S}$ for the condensate wave function $\psi$
reads:
\begin{equation}
{\mathcal S}=\int \d t {\mathcal
E}[\psi,\psi^*]-\frac{i\hbar}{2}\int \d t\d^3r \bigg(
\psi^*\frac{\partial \psi}{\partial t}-\psi\frac{\partial
\psi^*}{\partial t}
 \bigg),
\end{equation}
where the energy functional $\mathcal{E}$ is given by
\begin{equation}
{\mathcal E}[\psi,\psi^*]=\int \d^3r\bigg(
\frac{\hbar^2}{2m}|\mbox{$\boldmath{\nabla}$}\psi|^2+\frac{2\pi\hbar^2
a}{m}|\psi|^4+U({\bf r})|\psi|^2 \bigg).
\end{equation}

The fact that the transverse beam size is small compared with the
other relevant length scales allows us to consider the density
profile across the guide to have its equilibrium form appropriate
to the local number of particles per unit length. As a
consequence, the problem becomes essentially one dimensional. In
this case, the wave function can be assumed to have the form:
$\psi ({\bf r},t)=f(z,t)g(x,y;n)$ where $n$ is the local density
of particles per unit length and $g$ the equilibrium wave function
for the transverse motion. We emphasize that this ansatz does not
correspond to a complete separability between the axial and
transverse coordinates since $g$ depends on $z$ via the density
per unit length $n$. In the following, we choose $g$ to be
normalized to $\int |g|^2\d x\d y=1$, and therefore $|f|^2=n$.

The equations for $g$ and $f$ are obtained by minimization:
\begin{equation}
\frac{\delta}{\delta g^*}\bigg({\mathcal E}-\mu\int\d^3r
|\psi|^2\bigg)=0\qquad\mbox{and}\qquad \frac{\delta {\mathcal
S}}{\delta f^*}=0, \nonumber
\end{equation}
where $\mu$ is the chemical potential. After some algebra, one
obtains the following set of equations:
\begin{eqnarray}
&&\bigg(-\frac{\hbar^2}{2m}\Delta+U+\frac{4\pi\hbar^2a}{m}|fg|^2
\bigg)g=\tilde{\mu} g
\label{hd1}\\
&&i\hbar\frac{\partial f}{\partial
t}=-\frac{\hbar^2}{2m}\bigg(\frac{\partial^2 f}{\partial z^2}-
f\int |\mbox{$\boldmath{\nabla}$}g|^2\d x\d y\bigg)\nonumber\\&&+
\frac{4\pi\hbar^2a}{m}f|f|^2\int \d x\d y |g|^4+f\int \d x\d y
U|g|^2 \label{hd2}
\end{eqnarray}

where we introduce the effective chemical potential
$\tilde{\mu}=\mu-\hbar^2|\partial f/\partial z|^2/(2m|f|^2)$. To
obtain the hydrodynamic-like equations, it is convenient to write
$f$ in terms of a modulus and a phase: $f\equiv
\sqrt{n}e^{i\varphi}$. The velocity field $u$ associated with the
phase $\varphi$ reads $u\equiv (\hbar/m)\partial \varphi/\partial
z$. From (\ref{hd1}) and (\ref{hd2}), one derives the equations
for the density $n$ and the velocity $u$:
\begin{eqnarray}
&&  \frac{\partial n}{\partial t}+\frac{\partial }{\partial
z}(nu)=0
\\
&& m\frac{\partial u}{\partial t}=-\frac{\partial }{\partial z}
\bigg(
\tilde{\mu}+\frac{1}{2}mu^2-\frac{\hbar^2}{2m}\frac{1}{\sqrt{n}}\frac{\partial^2(\sqrt{n})}{\partial
z^2 }\bigg),\label{euler2}
\end{eqnarray}
The first equation is the usual equation of continuity, while the
second is an Euler-like equation. In the adiabatic limit, the last
term of (\ref{euler2}) is negligible.
\subsection{Determination of the chemical potential}
In the following, we derive the expression for the chemical
potential in two opposite limits, namely the Thomas Fermi regime
($na \gg 1$) and the weak coupling limit ($na \ll 1$). We restrict
our analysis to an harmonic trapping potential:
$U(x,y,z)=m\omega^2(x^2+y^2)/2$.
\subsubsection*{The Thomas-Fermi regime}
In this large density limit ($na \gg 1$), the kinetic term in
(\ref{hd1}) can be neglected. The normalization condition on $g$
provides the relation between the effective chemical potential
$\tilde{\mu}$ and the local density of particles per unit length:
$\tilde{\mu}=2\hbar\omega\sqrt{na}$. If $\omega$ does not depend
on $z$ or changes smoothly, $|\partial f/\partial z|\sim 0$ and
$\tilde{\mu}=\mu$, we therefore recover the results of Refs.
\cite{pethick98,pavloff}.
\subsubsection*{The weak-interaction limit}
In the absence of interaction, $g_0$ has a Gaussian form,
$|g_0|^2=(\pi\sigma^2)^{-1}e^{-(x^2+y^2)/\sigma^2}$ with
$\sigma=\sqrt{\hbar/m\omega}$ and $\mu_0=\hbar\omega$. The low
density limit corresponds to $na\ll 1$, we consequently derive
perturbatively from Eq. (\ref{hd1}) the correction to the chemical
potential:
\begin{equation}
\mu=\mu_0+\delta \mu=\mu_0+\frac{4\pi\hbar^2a}{m}n\int \d x\d
y|g_0|^2=\hbar\omega (1+2an).
\end{equation}


\begin{thebibliography}{00}
\bibitem{hdrq}
The set of hydrodynamic-like equations derived in Appendix A can
be recovered directly from hydrodynamic equations with the proper
choice for the ansatz. Actually, the radial dependence of the
velocity field has to be taken into account even if it does not
appear explicitly in the 1D equations.
\end{thebibliography}
\end{document}